\documentclass[prd,onecolumn,showpacs,floatfix,superscriptaddress,nofootinbib]{revtex4}
\usepackage[utf8]{inputenc}
\usepackage{graphicx}
\usepackage{epsfig}
\usepackage{bm}
\usepackage{amsfonts}
\usepackage[T1]{fontenc}
\usepackage{amssymb}
\usepackage{float}
\usepackage{amsmath}
\usepackage{dcolumn}
\usepackage{cancel}
\usepackage[colorlinks]{hyperref}
\usepackage[usenames,dvipsnames]{color}
\hypersetup{
     breaklinks=true,
    pdfstartview={FitH},    
    colorlinks=true,       
    linkcolor=blue,          
    citecolor=red,        
    filecolor=magenta,      
    urlcolor=blue,           
    anchorcolor=green,      
    linktocpage=true
}

\def\doi{http://doi.org}

\def\be{\begin{equation*}}
\def\ee{\end{equation*}}

\newcommand{\pd}{\partial}

\begin{document}

\title{Constraints on Scalar-Tensor Theory of Gravity by Solar System Tests }
\author{P. A. Gonz\'{a}lez}
\email{pablo.gonzalez@udp.cl} \affiliation{Facultad de
Ingenier\'{i}a y Ciencias, Universidad Diego Portales, Avenida Ej\'{e}rcito
Libertador 441, Casilla 298-V, Santiago, Chile.}
\author{Marco Olivares}
\email{marco.olivaresr@mail.udp.cl}
\affiliation{Facultad de
Ingenier\'{i}a y Ciencias, Universidad Diego Portales, Avenida Ej\'{e}rcito
Libertador 441, Casilla 298-V, Santiago, Chile.}
\author{Eleftherios Papantonopoulos}
\email{lpapa@central.ntua.gr}
\affiliation{Department of Physics, National Technical University of Athens, Zografou Campus GR 157 73, Athens, Greece.}
\author{Yerko V\'{a}squez}
\email{yvasquez@userena.cl}
\affiliation{Departamento de F\'{\i}sica, Facultad de Ciencias, Universidad de La Serena,\\
Avenida Cisternas 1200, La Serena, Chile.}
\date{\today}

\begin{abstract}

We study the motion of particles in the background of a scalar-tensor theory of gravity in which the scalar field is kinetically coupled to Einstein tensor. We constrain the value of the derivative parameter $z$  through solar system tests. By considering the perihelion precession we obtain the constrain $\sqrt{z}/m_p > 2.6\times 10^{12}$ m, the gravitational red-shift $\frac{\sqrt{z}}{m_{p}}>2.7\times10^{\,10}$ m, the deflection of light
 $\sqrt{z}/m_p > 1.6 \times 10^{11}$ m, and the gravitational time delay 	$\sqrt{z}/m_p > 7.9 \times 10^{12}$ m; thereby,
our results show that it is possible to constrain the value of the $z$ parameter in agreement with the observational tests that have been considered.



\end{abstract}

\maketitle


\tableofcontents

\newpage


\section{Introduction}

Modified theories of gravity were recently introduced in a attempt to cure certain inconsistencies of General Relativity (GR) and to explain certain observational results on dark matter and dark energy. These theories introduce modifications of GR on short and large distances in a attempt to provide
a viable theory of Gravity. The recent observational results on the Gravitational Waves (GWs) \cite{Abbott:2016blz,Abbott:2016nmj,Abbott:2017vtc,Abbott:2017oio,TheLIGOScientific:2017qsa} provide a new area of testing alternative gravity theories and differentiating them from GR. Therefore, it is very important to study   the compact objects predicted by different modified theories of gravity and possible GWs observational signatures  they might give. It would also be very interesting to study if classical solar system  tests on these objects such as light deflection, the perihelion shift of planets, the gravitational time-delay can give any discrepancy between the GR and the observations.

Some of the  simplest and extensively studied  modifications of GR are the scalar-tensor theories \cite{Fujii:2003pa}.  The presence of a scalar field coupled to gravity
results to black holes and compact objects dressed with a hairy matter distribution. The Horndeski Lagrangian
\cite{Horndeski:1974wa} provides one of the most well studied    scalar-tensor theories. This is because the   Horndeski theories
lead to second-order field equations and they result to
consistent theories without ghost instabilities \cite{Ostrogradsky:1850fid} and they preserve a classical
Galilean symmetry \cite{Nicolis:2008in,Deffayet:2009wt}. The Horndeski theory has been studied on short and large distances. Local black hole solutions were found in a subclass of Horndeski theories
which are characterized by the presence  of scalar field which is kinetically coupled to Einstein tensor \cite{Kolyvaris:2011fk,Rinaldi:2012vy,Kolyvaris:2013zfa,Babichev:2013cya,Charmousis:2014zaa}.
On large distances  the presence of the derivative coupling acts as a friction term in the inflationary period of the cosmological
evolution  \cite{Amendola:1993uh,Sushkov:2009hk,Germani:2010hd,Saridakis:2010mf,Huang:2014awa,Yang:2015pga,Koutsoumbas:2013boa}. This derivative coupling
introduces a mass scale in the theory which on large distances it can be constrained by the recent results on GWs.

If we assume that the dark energy drives the late cosmological expansion and it is parameterized by a  scalar field coupled to Einstein tensor it was  found \cite{Germani:2010gm,Germani:2011ua} that
the propagation speed of the tensor perturbations around the cosmological
Friedmann-Robertson-Walker (FRW) background is different from the speed of light $c$. Therefore the  measurement of the speed of GWs can be used
to constraint the value of the  derivative coupling and test in general  the applicability of the Horndeski theories at large distances
\cite{Lombriser:2015sxa,Lombriser:2016yzn,Bettoni:2016mij,Baker:2017hug,Creminelli:2017sry,Sakstein:2017xjx,Ezquiaga:2017ekz}.

The measurement of the speed of GWs by GW170817 and GRB170817A gave an upper bound on the speed of GWs $c_{gw}/c-1\le 7\times 10^{-16}$ \cite{Monitor:2017mdv}. If we assume that the peak of the GW signal and gamma ray burst GRB170817A were emitted simultaneously  then we get a lower bound $c_{gw}/c-1>-3\times 10^{-15}$ \cite{Monitor:2017mdv}, so we can safely conclude that $c_{gw}=c$. The precise measurement of the propagation speed of GWs is a very powerful tool to constraint the applicability of the Horndeski theory. Notably, in the Horndeski theory \cite{Horndeski:1974wa} and its generalizations \cite{Deffayet:2009mn}, the functions of the scalar field $\phi$ and its kinetic energy $X=-\pd_\mu\phi\pd^\mu\phi/2$, $G_4(\phi,X)$ and $G_5(\phi,X)$ should be constrained in order to be consistent with the above observations. The reason is that these terms provide the kinetic energy of the scalar field coupled to gravity and they influence the speed of GWs. The  term $G_5(\phi,X)$ is the general coupling of the scalar field to Einstein tensor and in  \cite{Gong:2017kim} assuming that the scalar field plays the role of dark energy   a lower bound on the mass scale present in this term was found  and combining the constraints from inflation the energy scale of the derivative coupling is bounded to be $10^{15}\text{GeV}\gg M \gtrsim 2\times 10^{-35}$GeV.

Modified gravity theories  can also be compared to GR predictions at relative small scales. Solar system observations, such as light deflection, the perihelion shift of planets, the gravitational time-delay among other are described within GR.   To study such effects you have to calculate the geodesics for the motion of particles around a black hole background. In  \cite{Chakraborty:2012sd} the perihelion precession of planetary orbits and the bending angle of null geodesics are estimated for different gravity theories in string-inspired models. The solar system effects have been studied in black hole AdS geometries by calculating the motion of particles on AdS spacetime \cite{Cruz:2004ts, Vasudevan:2005js, Hackmann:2008zz, Hackmann:2008zza, Olivares:2011xb, Cruz:2011yr,  Larranaga:2011fp, Villanueva:2013zta}. The motion of massless and massive particles in the background of  four-dimensional asymptotically AdS black holes with scalar hair \cite{Gonzalez:2013aca} were studied in \cite{Gonzalez:2015jna}. The geodesics are studied numerically and  the differences in the motion of particles between the four-dimensional asymptotically AdS black holes with scalar hair and their no-hair limit were discussed. In the context of solar system and astrophysical scenarios spherically symmetric solutions resulting from the coupling of the Gauss-Bonnet  with a scalar field were discussed in \cite{Bhattacharya:2016naa}.

Care should be taken when one studies specific scalar-tensor theories and compare their predictions with GR. In general scalar fields, depending on their coupling to gravity, mediate fifth forces. Therefore in these theories scalar fields should accommodate a mechanism to suppress the scalar interaction on small scales and make sure that precision tests of gravity at solar system scales are applicant.  There are various  screening mechanisms to suppress scalar interactions on small scales. One of the basic screening mechanism is the Vainshtein mechanism \cite{Vainshtein:1972sx} which was developed for the massive gravity (for an extensive review on the Vainshtein mechanism in massive gravity see \cite{Babichev:2013usa}).
The Vainshtein screening mechanism applies also to Galileon-like models~\cite{Nicolis:2008in} and to nonlinear massive gravity  \cite{nlmassive} in which
 the presence of nonlinear derivative scalar fields $\phi$'s
self-interactions can suppress the propagation of fifth forces
through the Vainshtein mechanism.
  In \cite{Kimura:2011dc,DeFelice:2011th} the consequences of the  Vainshtein mechanism was studied in scalar-tensor theories
taking into account the nonlinear effect. Therefore, for  models of modified gravity
we need to clarify the behavior of gravity
around and below the scale at which the relevant nonlinearities appear in order to test them against experiments and cosmological observations.

An extensive study of the  Vainshtein mechanism was carried out in \cite{Kase:2013uja} in the most general
scalar-tensor theories with second-order equations of motion resulted from a spherically
symmetric space-time with a matter source. They applied their general results to a number of concrete
models such as the covariant/extended Galileons
and the Dirac-Born-Infeld Galileons with Gauss-Bonnet and other terms.
They found that in these theories the fifth force can be suppressed
and be consistent to  solar system constraints,
provided that non-linear field kinetic terms coupled to
the Einstein tensor do not dominate
over other nonlinear field self-interactions.

The aim of this work is to constrain the parameters of the subclass of the Horndeski theory with a scalar field coupled kinetically to Einstein tensor
using the solar system observations. As we have already referred there are various black hole solutions in this subclass of the Horndeski theory. In \cite{Kolyvaris:2011fk} numerical black hole solutions were discussed with a scalar field coupled to Einstein tensor. Also in \cite{Babichev:2013cya}
static and spherically symmetric black hole solutions were found if the scalar field is time dependent. For our study we have to work with an exact black hole solution with a static matter distribution outside the horizon of the black hole. Another important requirement is that the derivative coupling of the  scalar field to Einstein tensor should appear in the metric function of the black hole in order to constrain this coupling from the solar system tests. A black hole solution that satisfies these constraints is the  well studied black hole solution of the Horndeski theory discussed in  \cite{Rinaldi:2012vy}. 

As we already discussed to apply the solar system test to our model  it should be consistent with GR.  For this to happen non-linear field kinetic terms should not dominate the dynamics. In \cite{Kase:2013uja} there is an extensive study of the Horndeski theories in which $G_5(\phi,X) \neq0$ as in the model we study. In the solution \cite{Rinaldi:2012vy} the coupling function of the kinetic scalar term to Einstein tensor is constant independent of the scalar field itself. Then it was shown in \cite{Kase:2013uja} that in this case we do not expect any nonlinearities to appear in our model and then GR can be recovered at small distances.    In this model because of the shift  symmetry the scalar field appears only through its derivative and then $\psi=\phi'$  appears as  an extra degree of freedom, expressed by the real quantity $\psi^2$. Following mainly the work in \cite{ Gonzalez:2018zuu,Olivares:2013jza} we will study the effects of the solar system tests by considering the perihelion precession, the gravitational red-shift, the deflection of light and the gravitational time delay.

The paper is organized as follows. In Section II we give a brief review of the four-dimensional Horndeski black hole of \cite{Rinaldi:2012vy} that we will consider as background. In Section III we study the motion of massless and massive particles, and we perform some classical tests such as the perihelion precession, the deflection of light and the gravitational time delay. Finally, in Section IV
we conclude.

\section{Four-Dimensional Horndeski Black Hole}
\label{bhHorn}

In this Section after reviewing the Horndeski theory we will discuss a particular hairy black hole solution \cite{Rinaldi:2012vy} of the Horndeski theory generated by a scalar field  non-minimally coupled to Einstein tensor.

The action of the Horndeski theory is given by \cite{Horndeski:1974wa},
\begin{equation}
\label{acth}
S=\int d^4x\sqrt{-g}(L_2+L_3+L_4+L_5)~,
\end{equation}
where
\begin{gather*}
L_2=K(\phi,X)~,\quad L_3=-G_3(\phi,X)\Box \phi~, \\ L_4=G_4(\phi,X)R+G_{4,X}\left[(\Box\phi)^2-(\nabla_\mu\nabla_\nu\phi)(\nabla^\mu\nabla^\nu\phi)\right]~, \\
L_5=G_5(\phi,X)G_{\mu\nu}\nabla^\mu\nabla^\nu\phi-\frac{1}{6}G_{5,X}[(\Box\phi)^3\\
-3(\Box\phi)(\nabla_\mu\nabla_\nu\phi)(\nabla^\mu\nabla^\nu\phi)+2(\nabla^\mu\nabla_\alpha\phi)(\nabla^\alpha\nabla_\beta\phi)(\nabla^\beta\nabla_\mu\phi)]~,
\end{gather*}
with $X=-\nabla_\mu\phi\nabla^\mu\phi/2$, $\Box\phi=\nabla_\mu\nabla^\mu\phi$,
the functions $K$, $G_3$, $G_4$ and $G_5$ are arbitrary functions of $\phi$ and $X$, and $G_{j,X}(\phi,X)=\partial G_j(\phi,X)/\partial X$ with $j=4,5$.

This action is the most general one for scalar-tensor theory with at most second-order field equations. If we take $K=G_3=G_5=0$ and $G_4=M_{\text{Pl}}/2$,
then we obtain Einstein's general relativity. If we take $G_3=G_5=0$, $K=X-V(\phi)$, and $G_4=f(\phi)$, then we get scalar-tensor $f(\phi) R$ theories.
If we take $G_4=M_{\text{Pl}}^2/2+X/(2M^2)$ or $G_4=M_{\text{Pl}}^2/2$ and $G_5=-\phi/(2M^2)$,
then we get the non-minimally derivative coupling $G_{\mu\nu}\nabla^\mu\phi\nabla^\nu\phi/(2M^2)$ with  the mass scale $M$.

As can be seen in the action (\ref{acth}) of the Horndeski theory, except the minimally coupling of the scalar field to gravity, there are other higher order couplings of the scalar field and also a term contained in the Lagrangian $L_5$ of the scalar field directly coupled to Einstein tensor. This term is interesting because it gives information of how strongly matter is coupled to curvature. Therefore, following the discussion we had in the introduction, it would be interesting to see how this coupling except the constrains it has from the GWs, it is further constrained from the solar system tests.

We will only  consider  the non-minimal derivative coupling of the scalar field to Einstein tensor of the Horndeski theory given by the Lagrangian
\begin{equation}
\mathfrak{L}=\frac{m_{p}^2}{2}R-\frac{1}{2}\left(g^{\mu\nu}-\frac{z}{m_{p}^2}G^{\mu\nu}\right)\partial_{\mu}\phi\partial_{\nu}\phi \,, \label{lagr}
\end{equation}
where $m_{p}$ is the Planck mass, $z$ is the derivative coupling of the scalar field to Einstein tensor, $G_{\mu\nu}$ the Einstein tensor,  $\varphi$ a scalar field, and $g_{\mu\nu}$ is the metric. The absence of scalar potential allows for the shift symmetry $\varphi\rightarrow \varphi+$const, which is the relevant Galileon symmetry that survives in curved space.

Consider the metric ansatz
\begin{equation}
\label{metric}
ds^2=-F(r)dt^2+G(r)dr^2+\rho ^2(r)(d\theta ^2+\sin^2{(\theta)} d\phi ^2)~.
\end{equation}

Setting $\rho=r$ the equations of motion are \cite{Rinaldi:2012vy}
\begin{align}
& r{F'\over F}={G-1}+{m_p^{2}r^{2}G\over z}+{Km_p^{2}G^{2}\over z \psi\sqrt{FG}}~, \label{current}\\
& r{F'\over F}={2m_p^{4}G(G-1)+z\psi^{2}(G-3)+m_p^{2}r^{2}G\psi^{2}\over (3z \psi^{2}+2m_p^{4}G)}~,\\
& {r\over 2}\left({F'\over F}-{G'\over G}\right)=  {2m_{p}G(G-1)-2z\psi^{2}-zr(\psi^2)'\over (3z \psi^{2}+2m_p^{4}G) }~,  \label{bern}
\end{align}
where $K$ is an integration constant and $\psi\equiv\varphi'$. We can see that if  $\psi=0$  implies $K=0$ and the resulting metric turns out to be the Schwarzschild one.  When $K=0$ and $z\neq 0$,  analytical  exact solutions of the system were found which they depend on the sign of $z$ and to avoid nonphysical modes for the scalar field   $z>0$ was considered
\begin{equation}
\label{lapsus}
F(r)=\frac{3}{4}+\frac{r^2 m_{p}^2}{12z}-\frac{2M}{m_{p}^2 r}+\frac{\sqrt{z}}{4m_{p} r}\arctan{\left(\frac{m_{p} r}{\sqrt{z}}\right)}\,,
\end{equation}
\begin{equation}
G(r)=\frac{(m_{p}^2 r^2 +2z)^2}{4(m_{p}^2 r^2+z)^2F(r)}\,,
\end{equation}
\begin{equation}
\psi^2(r)=-\frac{m_{p}^6 r^2(m_{p}^2 r^2 +2z)^2}{4z(m_{p}^2 r^2+z)^3F(r)}\,,
\end{equation}
 where $l^{2}=12z/m_{p}^{2}$ and $M$ is a constant of integration that will play the role of a mass. As it was discussed in \cite{Rinaldi:2012vy}  $z$ is a non-perturbative parameter when we regard the Lagrangian (\ref{lagr}) as a theory of modified gravity. Indeed, the deviation from GR vanishes when $z$ diverges and the scalar field is strongly coupled. Also, the parameter $z$ clearly interpolates between the flat black hole solution and the Schwarzschild AdS one as $1/z$ essentially plays the role of  an effective negative cosmological constant.

\section{Solar Test for the Horndeski Black Hole}
\label{STL}

In order to find the effects of the solar system to the Horndeski Black hole we have to study the geodesics of the space-time described by (\ref{metric}). For this we will solve the Euler-Lagrange equations for the variational problem associated with this metric.
The Lagrangian associated to the metric
(\ref{metric}) is given by
\begin{equation}\label{tl4}
  2\mathcal{L}=- F(r)\dot{t}^2+
 G(r)\dot{r}^2+r^2(\dot{\theta}^2+\sin^2\theta\,\dot{\phi}^2)=-m~,
\end{equation}
where $\dot{q}=dq/d\tau$, and $\tau$ is an affine parameter along the geodesic.
Since the Lagrangian (\ref{tl4}) is
independent of the cyclic coordinates ($t, \phi$), then their
conjugate momenta ($\Pi_t, \Pi_{\phi}$) are conserved and
the equations of motion read
\be \dot{\Pi}_{q} - \frac{\partial \mathcal{L}}{\partial q} = 0~,
\label{w.10} \ee
where $\Pi_{q} = \partial \mathcal{L}/\partial \dot{q}$
is the conjugate momenta of the coordinate $q$.  Using (\ref{tl4}), the above equation yields
\begin{equation}
\dot{\Pi}_{t} =0~, \quad \dot{\Pi}_{r} =-{\dot{t}^{2}\over 2}{dF(r)\over dr}+
{\dot{r}^{2}\over 2}{dG(r)\over dr}+r(\dot{\theta}%
^{2}+\sin^{2}\theta \dot\phi^2)~,
\label{w.11a}
\end{equation}
\begin{equation}
\dot{\Pi}_{\theta} = r^2\sin\theta \cos\theta \,\dot\phi^2, \quad
\textrm{and}\quad \dot{\Pi}_{\phi}=0~,
\label{w.11b}
\end{equation}
and the conjugate momenta are given by
\begin{equation}
\Pi_{t} = -F(r) \dot{t} , \quad \Pi_{r}= G(r)\dot{r}~,
\label{w.11c}
\end{equation}
\begin{equation}
\Pi_{\theta} = r^{2}\dot{\theta}~ , \quad
\textrm{and}\quad \Pi_{\phi}
= r^{2}\sin^{2}\theta \dot{\phi}~.
\label{w.11d}
\end{equation}
Now, without loss of generality, we consider that the motion develops in the invariant plane
 $\theta  = \pi/2$ and $\dot\theta =0$, which is characteristic of central fields. With this choice, Eqs. (\ref{w.11c}) and (\ref{w.11d})  become
 \begin{equation}
\Pi_{t} = -F(r) \dot{t}\equiv -\sqrt E~, \quad \Pi_{\phi}= r^{2}\dot{\phi}\equiv L~,
\label{w.11c2}
\end{equation}
where $E$ and $L$ are integration constants associated to each of them.
So, inserting
equations (\ref{w.11c2}) into equation (\ref{tl4}) we obtain
\begin{equation}\label{tl7}
  \left(\frac{dr}{d\tau}\right)^2=\frac{E-V(r)}{F(r)G(r)}~,
\end{equation}
where $V(r)$ is the effective potential
given by
\begin{equation}\label{tl8}
  V(r)=F(r)\left[m+\frac{L^2}{r^2}\right]~,
\end{equation}
where $m$ is the test mass, and by normalization we shall consider that $m=1$ for massive particles and $m=0$ for photons. Finally, using (\ref{w.11c2}) and (\ref{tl7}) we obtain the following equations
\begin{equation}\label{tl72}
  \left(\frac{dr}{dt}\right)^2=\frac{F(r)}{E}\left(\frac{E-V(r)}{G(r)}\right)~,
\end{equation}
\begin{equation}\label{tl73}
  \left(\frac{dr}{d\phi}\right)^2=\frac{r^4}{L^2}\left(\frac{E-V(r)}{F(r)G(r)}\right)~.
\end{equation}
\newline

In the following we will consider the regime: $r<\sqrt{z}/m_p$. Thus, for $0<m_pr/\sqrt{z}<1$
$\arctan m_pr/\sqrt{z} \approx \sum_{n=0}^{\infty }{(-1)^n\over 2n+1}(m_pr/\sqrt{z})^{2n+1}$. Therefore, the lapsus function Eq. (\ref{lapsus}) can be written as:

\begin{equation}
F(r)\approx \frac{3}{4}+\frac{r^2 m_{p}^2}{12z}-\frac{2M}{m_{p}^2 r}+\frac{\sqrt{z}}{4m_{p} r} \sum_{j=0}^{\infty }{(-1)^j\over 2j+1}
\left(\frac{m_{p} r}{\sqrt{z}}\right)^{2j+1}.
\end{equation}
Now, by considering the first three terms of the summation, we obtain
\begin{equation}\label{FF}
F(r)\approx 1-\frac{2M}{m_{p}^2 r}+\frac{r^4 m_{p}^4}{20z^2}\,,\quad
G(r)\approx \frac{1}{F(r)}\,.
\end{equation}
Note that, the lapse function approximates to Schwarzschild when $z\rightarrow \infty$, and $m_p=1$. With this approximation, the event horizon corresponds to the real solution of $F(r)=0$, given  by
\begin{equation}
r_+=\frac{2M}{m_p^2}
\, _4F_3\left[ \left\lbrace \frac{1}{5},\frac{2}{5},\frac{3}{5},\frac{4}{5}\right\rbrace
;\left\lbrace \frac{1}{2},\frac{3}{4},\frac{5}{4}\right\rbrace
;\frac{3125}{256} \left(-\frac{40 M z^2}{m_p\left(-20 z^2\right)^{5/4}}\right)^4\,\right]\,,
\end{equation}
where $_4F_3[\left\lbrace a_1,a_2,a_3,a_4 \right\rbrace,\left\lbrace b_1,b_2,b_3 \right\rbrace,x]$ is the generalized hypergeometric function.

\subsection{Time-like geodesics}
\begin{figure}[h]
\begin{center}
\includegraphics[width=0.38\textwidth]{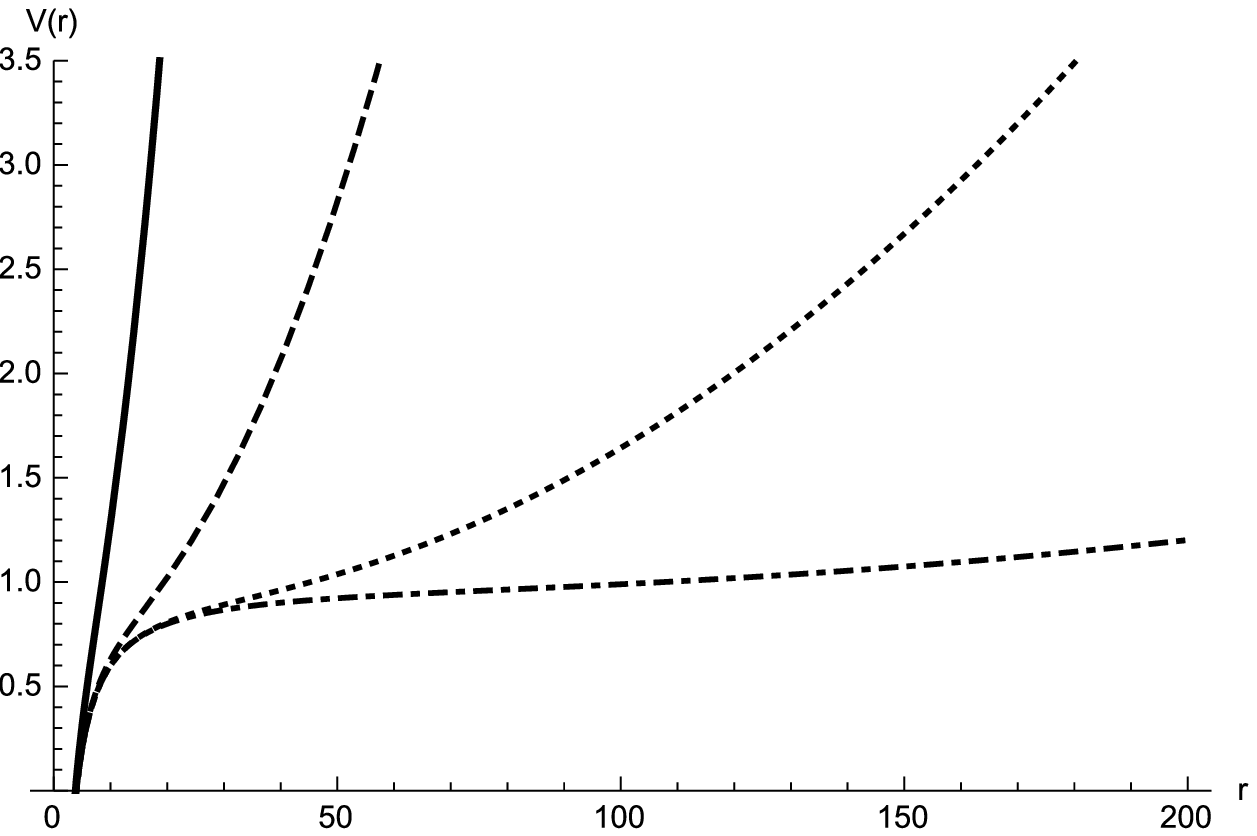}
\includegraphics[width=0.5\textwidth]{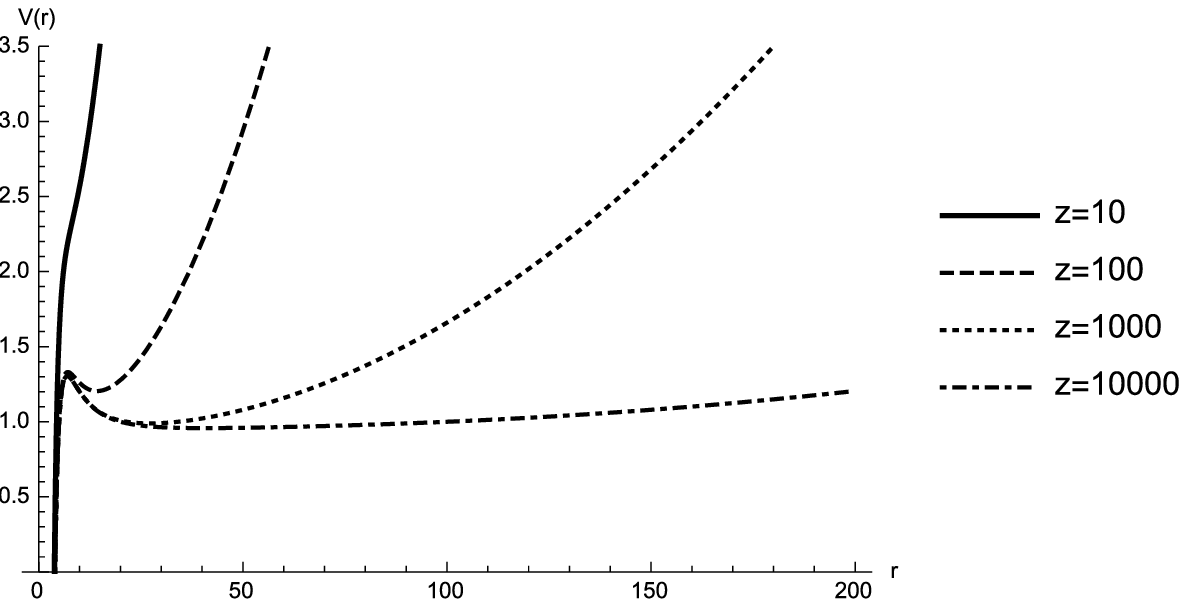}
\end{center}
\caption{The behavior of $V(r)$ for radial particles ($L=0$) left panel, and non-radial particles ($L=10$) right panel,  as a function of $r$, with $M=2$, and $m_p=1$.} \label{plots1}
\end{figure}

In order to observe the possible orbits, we plot the effective potential for massive particles (\ref{tl8}) which is shown in Fig. \ref{plots1}. In the following, we describe the radial motion and the angular motion.

\subsubsection{Radial motion}
\begin{figure}[h]
\begin{center}
\includegraphics[width=0.7\textwidth]{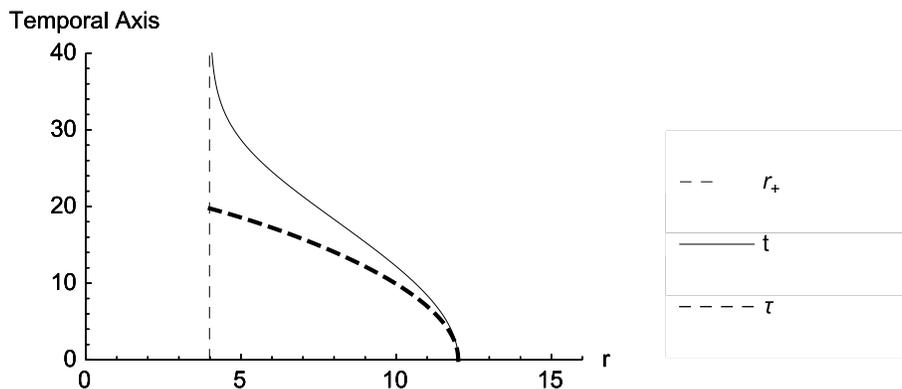}
\end{center}
\caption{The behavior of the proper ($\tau$) and the coordinate ($t$) time as a function of $r$, with $z=100$, $R_0=12$, and $V(R_0)=0.719$.} \label{plots2}
\end{figure}

In this case $L=0$. The particles always fall into the horizon from an upper distance $R_0$. Note that
the proper time ($\tau$) depends on the energy of the test particle, while that the coordinate time ($t$) does not depend on the energy of the test particle.
In Fig. \ref{plots2} we plot the proper time ($\tau$) and the coordinate time ($t$) as a function of $r$ for a particle falling from a finite distance with zero initial velocity, we can see that the particle falls towards the horizon in a finite proper time. The situation is very different if we consider the trajectory in the coordinate time, where $t$ goes to infinity. This physical result is consistent with the Schwarzschild  black hole.

\subsubsection{Angular motion}
For the angular motion we consider $L>0$. The allowed orbits depend on the value of the constant $E$.

\begin{figure}[h]
\begin{center}
\includegraphics[width=0.5\textwidth]{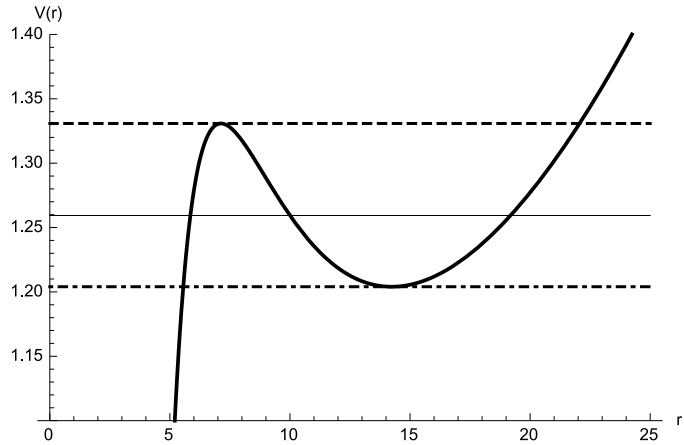}
\end{center}
\caption{The behavior of $V(r)$ for non-radial particles ($L=10$) as a function of $r$, with $M=2$, $z=100$, and $m_p=1$. The dashed line corresponds to $E=E_u\approx1.331$, the dot dashed line corresponds to $E=E_s\approx 1.204$ and the thin line corresponds to $E=1.259$.} \label{plots3}
\end{figure}

\begin{itemize}
\item
If $E=E_s\approx 1.204$ the particle can orbit in a stable
circular orbit at $r_s=14.230$, see Fig. \ref{plots3}.

\item
If $E=E_u\approx1.331$ the particle can orbit in an unstable
circular orbit at $r_u=7.126$. Also, there are two critical orbits that approximate asymptotically to the unstable circular orbit. For the first kind orbit the particle starts from rest and at a finite distance greater than the unstable radius. For the second kind orbit the particle starts from a finite distance greater than the horizon, but smaller than the unstable radius.

\item
The planetary orbits are constrained to oscillate between
an apastron and a periastron. We plot in Fig. \ref{plot4}
the planetary orbit for $E=1.259$. We can observe that the particle completes an oscillation in an angle greater than $2\pi$ which is similar to the Schwarzschild black hole \cite{chandra}.
\begin{figure}[h]
\begin{center}
\includegraphics[width=0.4\textwidth]{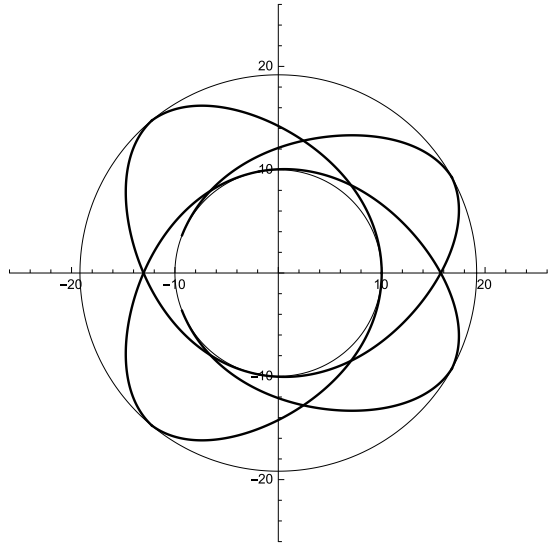}
\end{center}
\caption{The planetary orbit  for $L=10$, $r_A=19.185$, $r_P=10$ and $E=1.259$.} \label{plot4}
\end{figure}

\end{itemize}

It is possible to calculate the periods of the circular orbits ($r_{c.o.}$), which can be stable ($r_s$) or unstable ($r_u$) orbits using the constants of motion $\sqrt{E}$ and $L$, given by (\ref{w.11c2}), which yields
\begin{equation}
T_{\tau}=2\pi \sqrt{\frac{2r_{c.o.}F(r_{c.o.})-r_{c.o.}^2F'(r_{c.o.})}{F'(r_{c.o.)}}}~,
\end{equation}
and
\begin{equation}
T_t=2\pi \sqrt{\frac{2r_{c.o.}}{F'(r_{c.o.})}}~,
\end{equation}
where $T_{\tau}$ is the period of the orbit with respect to the proper time and $T_t$ is the period of the orbit with respect to the coordinate time. It is worth to mention that the periods depend on the lapsus function and their derivative that contain the $M$ and $z$ parameter. On the other hand, for the stable circular orbits is possible to find the epicycle frequency, given by $\kappa^2=V''(r_s)/2$, which yields
\begin{equation}
    \kappa^2=\frac{F''(r_s)}{2}+\frac{F'(r_s)}{2}\left( \frac{r_s^2F''(r_s)-4r_sF'(r_s)+6F(r_s)}{2r_sF(r_s)-r_s^2F'(r_s)}\right)\,.
\end{equation}

\subsubsection{Perihelion precession}

The following treatment, performed by Cornbleet \cite{Cornbleet},
allows us to derive the formula for the advance of the
perihelia of planetary orbits.
The starting point is to
consider the line element in unperturbed Lorentz coordinates
\begin{equation}
ds^2 = -dt^2 + dr^2 + r^2(d\theta^2 + sin^2 \theta d\phi^2)~,
\end{equation}
together with line element (\ref{metric}), where $F(r)$ and $G(r)$ are given by (\ref{FF}). So, considering
only the radial and time coordinates in the binomial
approximation, the transformation gives
\begin{equation}
d\tilde{t}\approx \left(1-\frac{M}{m_{p}^2 r}+\frac{r^4 m_{p}^4}{40z^2}\right) dt~,
\end{equation}
\begin{equation}\label{r}
d\tilde{r}\approx \left(1+\frac{M}{m_{p}^2 r}-\frac{r^4 m_{p}^4}{40z^2}\right) dr~.
\end{equation}
We will consider two elliptical orbits: the classical
Kepler orbit in $(r, t)$ space and a Horndeski orbit in an
$(\tilde{r},\tilde{t})$ space. Then, in the Lorentz space $dA = \int_0^{\mathcal{R}}rdrd\phi=\mathcal{R}^2d\phi/2$, and hence
\begin{equation}
\frac{dA}{dt}=\frac{1}{2}\mathcal{R}^2\frac{d\phi}{dt}~,
\end{equation}
which corresponds to Kepler's second law. For the
Horndeski case we have
\begin{equation}\label{A}
d\tilde{A}=\int_0^{\mathcal{R}}rd\tilde{r}d\phi~,
\end{equation}
where $d\tilde{r}$ is given by Eq. (\ref{r}).
So, we can write (\ref{A}) as
\begin{eqnarray}\label{A2}
\nonumber d\tilde{A}&=&\int_0^{\mathcal{R}}r\left(1+\frac{M}{m_{p}^2 r}-\frac{r^4 m_{p}^4}{40z^2}\right) drd\phi\\
&&\approx \frac{\mathcal{R}^2}{2}\left(1+\frac{2M}{m_{p}^2 \mathcal{R}}-\frac{\mathcal{R}^4 m_{p}^4}{120z^2}\right)d\phi  \,~.
\end{eqnarray}
Therefore, applying the binomial approximation wherever
necessary, we obtain
\begin{eqnarray}
\nonumber \frac{ d\tilde{A}}{d\tilde{t}}&=&\frac{\mathcal{R}^2}{2}\left(1+\frac{2M}{m_{p}^2 \mathcal{R}}-\frac{\mathcal{R}^4 m_{p}^4}{120z^2}\right)\frac{d\phi}{d\tilde{t}}\\
&&\approx \frac{\mathcal{R}^2}{2}\left(1+\frac{2M}{m_{p}^2 \mathcal{R}}-\frac{\mathcal{R}^4 m_{p}^4}{120z^2}\right) \left(1+\frac{M}{m_{p}^2 \mathcal{R}}-\frac{\mathcal{R}^4 m_{p}^4}{40z^2}\right) \frac{d\phi}{dt}\,~.
\end{eqnarray}
So, using this increase to improve the elemental angle
from $d\phi$ to $d\tilde{\phi}$, then for a single orbit
\begin{equation}\label{so}
\int_0^{\Delta\tilde{\phi}}d\tilde{\phi}=\int_0^{\Delta\phi=2\pi}\left(1+\frac{3M}{m_{p}^2 \mathcal{R}}-\frac{\mathcal{R}^4 m_{p}^4}{30z^2} \right)d\phi\,~,
\end{equation}
where we have neglected products of $M$ and $z$.
The polar form of an ellipse is given by
\begin{equation}\label{ell}
\mathcal{R}=\frac{l}{1+\epsilon \, \cos\phi}  \,~,
\end{equation}
where $\epsilon$ is the eccentricity and $l$ is the semi-latus rectum.
In this way, plugging Eq. (\ref{ell}) into Eq. (\ref{so}), we obtain
\begin{eqnarray}
\Delta\tilde{\phi}=2\pi+\frac{3M}{m_{p}^2}\int_0^{2\pi} \frac{1+\epsilon \,cos\phi }{l} d\phi-\frac{ m_{p}^4}{30z^2} \int_0^{2\pi} \left(\frac{l}{1+\epsilon \,\cos\phi } \right)^4 d\phi\,~,
\end{eqnarray}
which at first order yields
\begin{eqnarray}
\Delta\tilde{\phi}\approx 2\pi+\frac{6\pi M}{m_{p}^2\,l}-\frac{ \pi m_{p}^4\,l^4}{15z^2}\,~.
\end{eqnarray}

Note that if we consider the limit ${M\over m_p^2} 	\rightarrow M_{\odot}$ and $z 	\rightarrow  \infty$, we recovered the classical result for the Schwarzschild spacetime. Therefore, the perihelion advance has the standard value
of GR plus the correction term coming from the Horndeski's theory.
It is worth to mention that
		the observational value of the precession of perihelion for Mercury is
		$\Delta\tilde{\phi}_{Obs.}=5599.74$,(arcsec/Julian-century) \cite{observado}, and the total is
		$\Delta\tilde{\phi}_{Total}= 5603.24$,(arcsec/Julian-century) \cite{total, Will}, with a difference between them of $\Delta\tilde{\phi}=-3.50$,(arcsec/Julian-century), which is possible to be attributed as a correction coming from  a scalar-tensor theory, in particular coming from the  parameter $z$ of the Horndeski's theory ($\Delta\tilde{\phi}=-\frac{ \pi m_{p}^4\,l^4}{15z^2} $), giving the constrain
	$\sqrt{z}/m_p \geq 2.6\times 10^{12}(m)$ that allow us a better accuracy between the observational value and the theoretical value of the precession of perihelion for Mercury.


\subsection{Null geodesics}

In the next analysis, we consider two kinds of motion: radial motion ($L=0$) and angular motion ($L>0$) of the photons ($m=0$).

\subsubsection{Radial motion}

\begin{figure}[h]
\begin{center}
\includegraphics[width=0.6\textwidth]{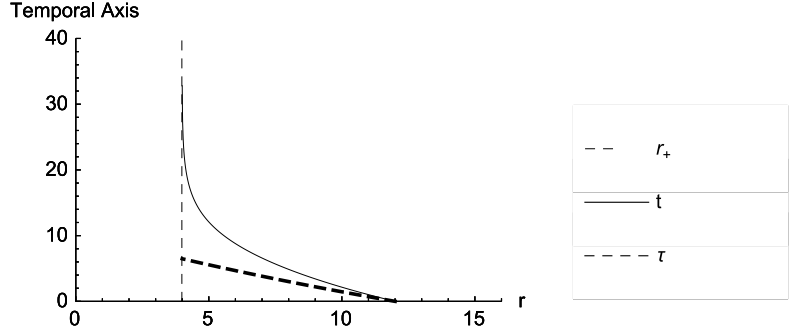}
\end{center}
\caption{The behavior of the affine ($\tau$) and the coordinate ($t$) time as a function of $r$, with $z=100$ and $E=1$.}
\label{plot5}
\end{figure}

In this case, the master equation (\ref{tl7}) can be written as
\begin{equation}
 \frac{dr}{d\tau}=\pm \sqrt{E}~,
\end{equation}
where ($+$) stands for outgoing photons and $(-)$ stands for ingoing photons. The solution of the above equation yields
\begin{equation}
r=\pm \sqrt{E} \tau + r_0~,
\end{equation}
where $r_0$ is an integration constant that corresponds to the initial position of the photon, as in the Schwarzschild case.  The photons always fall into the horizon from an upper distance.
In Fig. \ref{plot5} we plot the affine  ($\tau$) and coordinate ($t$) time as a function of $r$ for a photon falling from a finite distance ($r_0=12$), we can see that photons fall towards the horizon in a finite affine time. The situation is very different if we consider the trajectory in the coordinate time, where $t$ goes to infinity.

\subsubsection{Angular motion}
\begin{figure}[h]
\begin{center}
\includegraphics[width=0.6\textwidth]{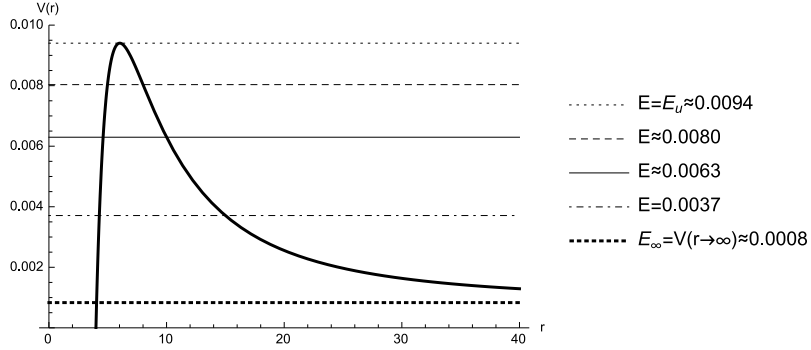}
\end{center}
\caption{The behavior of the effective potential for photons as a function of $r$, for $L=1$, $M=2$, $m_p=1$, and $z=100$.}
\label{plot6}
\end{figure}

In this case, the allowed orbits for photons depend on the value of the
impact parameter $b\equiv L/\sqrt{E}$. Next, based on the impact parameter values shown in Fig. \ref{plot6}, where $E_u$ is the energy of the unstable circular orbit and $E_\infty=V(r\rightarrow \infty) = \frac{L^2m_p^2}{12z}$,
 we give a brief qualitative description of the allowed
angular motions for photons,
described in the following

\begin{figure}[h]
\begin{center}
\includegraphics[width=0.3\textwidth]{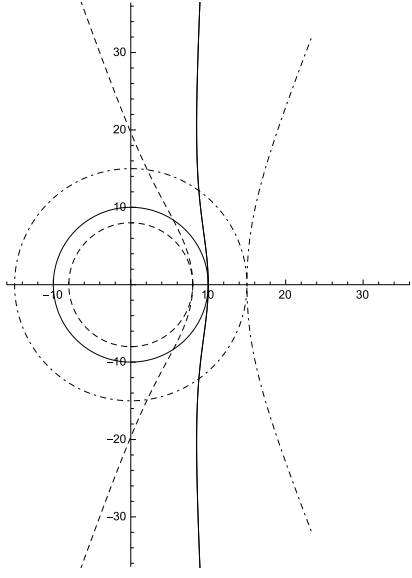}
\end{center}
\caption{The deflection of light for $L=1$ and $z=100$. The dashed line corresponds to $E=0.0080$ with $r_D=8$, the continuous line corresponds to $E=0.0063$ with $r_D=10$ and the dot-dashed line corresponds to $E=0.0037$ with $r_D=15$.} \label{plot8}
\end{figure}

\begin{itemize}
  \item \emph{Capture zone}:
  If $0< b < b_{u}$, photons fall inexorably
  to the horizon,
  and their cross section,
  $\sigma$, in this geometry is \cite{wald}
  \begin{equation}\label{mr51}
    \sigma=\pi\,b_u^2~.
  \end{equation}
\item \emph{Critical trajectories}:
If $b=b_{u}$ ($E_u \approx 0.009$), photons can stay in one of the unstable
  inner circular orbit of radius  $r_{u}$ ($r_u\approx 6.03$).
  Therefore, photons that arrive from the initial distance
  $r_i$ ($r_+ < r_i< r_u$, or $r_u< r_i<\infty$)
  can fall asymptotically into a circle of radius $r_{u}$~.
  The period respect to the affine parameter ($\tau$) for the unstable circular orbit is
  \begin{equation}\label{p1}
  T_{\tau}=\frac{2\pi r_u^2}{L}\,.
  \end{equation}
  Also, the coordinate period is given by
  \begin{equation}\label{p2}
  T_t=2\pi b_u\,.
  \end{equation}
 \item \emph{Deflection zone}. If $b_{u} <b <b_{0}\equiv L/\sqrt{E_\infty} $, photons can fall from
infinity to a minimum distance $r_D$ and return to infinity. This photons are deflected, see Fig. \ref{plot8}. Also, we can observe a zone where the deflection is attractive and other one repulsive. The
other allowed orbits correspond to photons moving into the
other side of the potential barrier, which plunges into the
singularity.
\end{itemize}

\subsubsection{Gravitational redshift}

Since Horndeski black hole is a stationary spacetime there is a time-like Killing vector so that in coordinates adapted to the symmetry the ratio of the measured frequency of a light ray crossing different positions is given by
\begin{equation}
{\nu \over \nu_0}=\sqrt{\frac{g_{00}(r)}{g_{00}(r_0)}}~,
\end{equation}
for  $M/(m_p^2r)<<1$ and $rm_p/\sqrt{z}<<1$, we obtain
\begin{equation}
{\nu \over \nu_0}\approx  1+\frac{M}{m_{p}^2 r_0}-\frac{M}{m_{p}^2 r}
+\frac{r^4 m_{p}^4}{40z^2} -\frac{r_0^4 m_{p}^4}{40z^2}\,~,
\end{equation}
where we neglected products of $z$ and $M$. Obviously, if we consider the limit ${M\over m_p^2} 	\rightarrow M_{\oplus}$ and $z 	\rightarrow  \infty$, we recover the classical result for the Schwarzschild spacetime.
The clock can be compared with an accuracy of $10^{-15}$, the H-maser in the GP-A redshift experiment \cite{Vessot:1980zz} reached an accuracy of $10^{-14}$. Therefore, by considering that all observations are well described within Einstein's theory, we conclude that the extra terms of Horndeski must be $<10^{-14}$. Thus,
\begin{equation}
\frac{\sqrt{z}}{m_{p}}>2.7\times10^{\,10}\, m\,,
\end{equation}
where we assume a clock comparison between Earth and a satellite at 15,000 km height, as in Ref. \cite{Kagramanova:2006ax}.

\subsubsection{Deflection of light}

Currently, the light-deflection phenomenon is an important phenomena in the modern astronomy by its application to the gravitational lenses. In order to study the bending of the light, we consider Eq. (\ref{tl73}) for photons, which can be written as
\begin{equation}
\left(\frac{dr}{d\phi}\right)^2=\frac{r^{4}}{b^2}- r^2+{2M\over m_p^2} r-{ m_p^4\over 20z^2} r^6\,,
\end{equation}
where $b$ is the impact parameter, and we have used Eq. (\ref{FF}). Now, by performing the change of variables $r=1/u$, the above equation yields
\begin{equation}\label{ue}
\left(\frac{du}{d\phi}\right)^2=\frac{1}{b^2}- u^2+{2M\over m_p^2} u^3-{ m_p^4\over 20z^2}\, u^{-2}\,.
\end{equation}
Notice that for $z \rightarrow \infty$, and $m_p=1$, the above equation is reduced to the classical equation of Schwarzschild for the motion of photons given by
\begin{equation}
\left(\frac{du}{d\phi}\right)^2=\frac{1}{b^2}-u^2+2M u^{3}\,~.
\end{equation}
So, the derivative of Eq. (\ref{ue}) with respect to $\phi$ yields
\begin{equation}
u^{\prime\prime}+ u ={3M\over m_p^2} u^2+{ m_p^4\over 20z^2}\, u^{-3}\,,
\end{equation}
where ${}^\prime$ denotes the derivative with respect to $\phi$. So, following the procedure established in Ref. \cite{Straumann}, we obtain
\begin{equation}
u=\frac{1}{b}\sin(\phi)+{ 3M\over 2m_p^2b^2}+{ m_p^4b^3\sqrt{2}\over 10z^2}
+\left({ M\over 2m_p^2b^2}-{ m_p^4b^3\sqrt{2}\over 20z^2}\right)\cos(2\phi)\,~.
\end{equation}

In the limit $u \rightarrow 0$, $\phi$ approaches $\phi_{\infty}$, with
\begin{equation}
-\phi_{\infty}={ 2M\over m_p^2b}+{ m_p^4b^4\sqrt{2}\over 20z^2}\,~.
\end{equation}
Therefore, for the Horndeski black holes the deflection of light $\hat{\alpha}$ is equal to $2\left |-\phi_{\infty}\right |$ and yields
\begin{equation}\label{GB1}
\hat{\alpha}={4M\over m_p^2b}+{ m_p^4b^4\sqrt{2}\over 10z^2}\,~.
\end{equation}

Note that if we consider the limit ${M\over m_p^2} 	\rightarrow M_{\odot}$ and $z 	\rightarrow  \infty$, we recovered the classical result of GR; that is, $\hat{\alpha}_{GR}=4M_{\odot}/b$. If the impact parameter is equal to the radius of the sun, the value obtained is $\hat{\alpha}_{GR}=4M_{\odot}/R_{\odot}=1.75092''$.
 The parameterized post-Newtonian (PPN) formalism introduces the phenomenological parameter $\gamma$, which characterizes the contribution of space curvature to gravitational deflection. In this formalism the deflection angle is $\hat{\alpha}= 0.5(1 + \gamma)1.7426$, and currently  $\gamma= 0.9998 \pm 0.0004$  \cite{Shapiro}. So,  $\hat{\alpha}=  1.74277''$ for $\gamma= 0.9998 + 0.0004$  and $\hat{\alpha}=1.74208''$ for $\gamma= 0.9998 - 0.0004$.
 The observational values, compared to the classic result, are smaller, and the contribution of the Horndeski term to the deflection angle is positive, therefore, there is no observable effect. Thus, if the Horndeski term contributes it does so that $\hat{\alpha}_{Horndeski}< 0.00001'' $, or
 $\sqrt{z}/m_p > 1.6 \times 10^{11}(m)$.

\subsubsection{Gravitational time delay}

An interesting relativistic effect in the propagation of light rays is the apparent delay in the time of propagation for a light signal passing near the Sun, which is a relevant correction for astronomic observations, and is called the Shapiro time delay. The time delay of Radar Echoes corresponds to the determination of the time delay of radar signals which are transmitted from the Earth through a region near the Sun to another planet or  spacecraft and then reflected back to the Earth. The time interval between emission and return of a pulse as measured by a clock on the Earth is

\begin{equation}
t_{12}=2\, t(r_1,\rho_0)+2\, t(r_2,\rho_0)~,
\end{equation}
where $\rho_0$ as closest approach to the Sun. Now, in order to calculate the time delay we use (\ref{tl72}), (\ref{FF}) and the coordinate time
\begin{equation}
\dot{r}=\dot{t}\,\frac{dr}{dt}=\frac{E}{F(r)}\frac{dr}{dt}\,~,
\end{equation}
so, (\ref{tl7}) can be written as
\begin{equation}\label{ct}
\frac{E}{F(r)}\frac{dr}{dt}=\sqrt{E^2-\frac{L^2}{r^2}F(r)}\,~.
\end{equation}
By considering $\rho_0$ as closest approach to the Sun, $dr/dt$ vanishes, so that
\begin{equation}\label{TD1}
\frac{E^2}{L^2}=\frac{F(\rho_0)}{\rho_0^2}~.
\end{equation}
Now, by inserting (\ref{TD1}) in (\ref{ct}), the coordinate time which the light requires to go from $\rho_0$ to $r$ is
\begin{equation}
t(r,\rho_0)=\int_{\rho_0}^r \frac{dr}{F(r)\sqrt{1-\frac{\rho_0^2}{F(\rho_0)}\frac{F(r)}{r^2}}}~.
\end{equation}
So, at first order correction we obtain
\begin{eqnarray}
\nonumber t(r, \rho_0)&=&\sqrt{r^2-\rho_0^2}+\frac{M}{m_p^2}\left[\sqrt{\frac{r-\rho_0}{r+\rho_0}}+2ln\left(\frac{r+\sqrt{r^2-\rho_0^2}}{\rho_0}\right)\right]+
\\
&&+\frac{m_p^4}{300z^2}\sqrt{r^2-\rho_0^2}\left[\frac{9\rho_0^2}{2} -\frac{3\rho_0^2r^2}{2}-3r^4 \right]~.
\end{eqnarray}
Therefore, for the circuit from point 1 to point 2 and back the delay in the coordinate time is
\begin{equation}
\Delta t := 2\left[t(r_1, \rho_0)+t(r_2,\rho_0)-\sqrt{r_1^2-\rho_0^2}-\sqrt{r_2^2-\rho_0^2}\right]=\Delta t_M+\Delta t_z~,
\end{equation}
where
\begin{eqnarray}
\Delta t_M&=&\frac{2M}{m_p^2}\left[ 2\, ln\left(\frac{(r_1+\sqrt{r_1^2-\rho_0^2})(r_2+\sqrt{r_2^2-\rho_0^2})}{\rho_0^2}\right)+\sqrt{\frac{r_1-\rho_0}{r_1+\rho_0}}
+\sqrt{\frac{r_2-\rho_0}{r_2+\rho_0}}\right]~,\\
\Delta t_z&=&\frac{m_p^4}{300z^2}\left[ \sqrt{r_1^2-\rho_0^2}\left( \frac{9\rho_0^2}{2} -\frac{3\rho_0^2r_1^2}{2}-3r_1^4 \right)  +\sqrt{r_2^2-\rho_0^2}\left( \frac{9\rho_0^2}{2} -\frac{3\rho_0^2r_2^2}{2}-3r_2^4 \right) \right]~.
\end{eqnarray}

For a round trip in the solar system, we have ($\rho_0<<r_1,r_2$)
\begin{equation}
\Delta t \approx \frac{4M}{m_p^2}\left[ 1+ ln\left(\frac{4r_1r_2}{\rho_0^2}\right)\right]-\frac{m_p^4}{100z^2}\left(r_1^5+r_2^5\right)=\Delta t_{GR}+\Delta t_{Horndeski}~.
\end{equation}

 Note that if we consider the limit ${M\over m_p^2} 	\rightarrow M_{\odot}$ and $z 	\rightarrow  \infty$, we recover the classical result of GR; that is, $\Delta t_{GR}=4M_{\odot}\left[ 1+ ln\left(\frac{4r_1r_2}{\rho_0^2}\right)\right]$. For a round trip from the Earth to Mars and back, we get (for $\rho_0 \ll r_1 , r_2$ ), where $r_1 \approx r_2=2.25\times 10^{11} m$  is the average distance Earth-Mars. Considering $\rho_0$, as closest approach to the Sun, like the radius  of the Sun ($R_{\odot} \approx 
 6.960\times 10^{8}\,m$) plus the solar corona ($  \sim 10^{9}m$), $\rho_0 \approx1.696\times 10^{9}m$, then,  the time delay is $\Delta t_{GR} \approx 240\, \mu\,s$.  To give an idea of the experimental possibilities, we mention that the error in the time measurement of a circuit during the Viking mission was only about $10\, ns$ \cite{Straumann}. If the Horndeski term contributes, $\Delta t_{Horndeski}=-\frac{m_p^4}{100z^2}\left(r_1^5+r_2^5\right)$, it does so that $\Delta t_{Horndeski}< 10^{-8}\,s$, or
	$\sqrt{z}/m_p > 7.9 \times 10^{12}(m)$. \\


\section{Concluding comments}
\label{sec:conclution}

We considered four-dimensional Horndeski black holes and we analyzed the motion of particles in these background with the objective to study the geodesics and to constrain the value of the derivative coupling  parameter $z$ through solar system tests.

Concerning  the radial and angular time-like geodesics,  we found that the motion of particles is confined, while, for the Schwarzschild spacetime the radial and angular time-like geodesics are  confined but there are also unconfined geodesics. However, both spacetimes allows the existence of stable and unstable circular orbits, as well as, the existence of planetary orbits. For null geodesics, four-dimensional Horndeski and Schwarzschild spacetimes allows the existence of a capture zone, an unstable circular orbit, and a deflection zone. However, for Schwarzschild spacetime the effective potential vanishes at infinity, while for Horndeski spacetime the effective potential go to $\frac{L^2mp^2}{12z}$ at infinity; thus, for $z\rightarrow \infty$ we recover GR. In this way, the behavior of the geodesics is qualitatively similar to the behavior of the geodesics in a Schwarzschild AdS spacetime \cite{Cruz:2004ts}.

In respect to constrain the value of the $z$ parameter through solar system tests, we considered the perihelion precession, the gravitational red-shift, the deflection of light and the gravitational time delay. Our results show that it is  possible to constrain the value of the derivative coupling  parameter  $z$ ($\sqrt{z}/m_p > 2.6\times 10^{12}$ m) in agreement with all the observational tests that have been considered.

We found that the perihelion advance has the standard value
of GR plus the correction term coming from the Horndeski's theory, given by $\Delta\tilde{\phi}\approx 2\pi+\frac{6\pi M}{m_{p}^2\,l}-\frac{ \pi m_{p}^4\,l^4}{15z^2}$; thus, we  constrained the $z$ parameter to $\sqrt{z}/m_p\geq 2.6\times 10^{12}$ m, in order to obtain a better accuracy between the observational value and the theoretical value of the precession of perihelion for Mercury.

Also, we obtained that the gravitational red-shift is ${\nu \over \nu_0}\approx  1+\frac{M}{m_{p}^2 r_0}-\frac{M}{m_{p}^2 r}
+\frac{r^4 m_{p}^4}{40z^2} -\frac{r_0^4 m_{p}^4}{40z^2}$, and in the limit ${M\over m_p^2} 	\rightarrow M_{\oplus}$ and $z 	\rightarrow  \infty$, the classical result for the Schwarzschild spacetime is recovered; thus, by considering that this observation is well described within Einstein's theory, the extra terms of Horndeski must be $<10^{-14}$, which implies the constraint   $\frac{\sqrt{z}}{m_{p}}>2.7\times10^{\,10}$ m.

The deflection of light is given by $\hat{\alpha}={4M\over m_p^2b}+{ m_p^4b^4\sqrt{2}\over 10z^2}$, and  in the limit ${M\over m_p^2} 	\rightarrow M_{\odot}$ and $z 	\rightarrow  \infty$, we recovered the classical result of GR; that is, $\hat{\alpha}_{GR}=4M_{\odot}/b$. In this case, we shown that there is a  zone where the deflection is attractive and another zone where the deflection is  repulsive. Also, the observational values compared to the classic result are smaller, and the contribution of the Horndeski term to the deflection angle is positive; therefore, there is no observable effect. Thus, if the Horndeski term contributes it does so that $\hat{\alpha}_{Horndeski}< 0.00001'' $, or $\sqrt{z}/m_p > 1.6 \times 10^{11}$ m.

Finally, the gravitational time delay for a round trip in the solar system, we found that
$\Delta t  =\Delta t_{GR}+\Delta t_{Horndeski}$, where $\Delta t_{GR}=4M_{\odot}\left[ 1+ ln\left(\frac{4r_1r_2}{\rho_0^2}\right)\right]$, and $\Delta t_{Horndeski}=-\frac{m_p^4}{100z^2}\left(r_1^5+r_2^5\right)$; thus, in the limit ${M\over m_p^2} 	\rightarrow M_{\odot}$ and $z 	\rightarrow  \infty$, we recover the classical result of GR, for a round trip from the Earth to Mars and back,  the time delay is $\Delta t_{GR} \approx 240\, \mu\,s$. However, the error in the time measurement of a circuit during the Viking mission was only about $10\, ns$ \cite{Straumann}. Therefore,  if the Horndeski term contributes, it does so for $\Delta t_{Horndeski}< 10^{-8}\,s$, or $\sqrt{z}/m_p > 7.9 \times 10^{12}$ m.



\acknowledgments

We thank the referee for his/her careful review of the manuscript and his/her valuable comments and suggestions which helped us to improve the manuscript. We thank Eugeny Babichev and Shinji Tsujikawa for their valuable comments and remarks. Y.V. acknowledge support by the Direcci\'on de Investigaci\'on y Desarrollo de la Universidad de La Serena, Grant No. PR18142.

\end{document}